\documentclass[aps,prl,twocolumn,floats,showpacs]{revtex4}
\usepackage{epsfig}
\usepackage{graphicx}
\usepackage{amsmath}

\newcommand{\beq}{\begin{equation}}
\newcommand{\eeq}{\end{equation}}
\newcommand{\bea}{\begin{eqnarray}}
\newcommand{\eea}{\end{eqnarray}}

\begin{document}
\title{A spin-boson thermal rectifier}
\author{\footnote{Present address: Department of Chemical Physics, The Weizmann Institute of Science, Rehovot 76100, Israel}Dvira Segal and Abraham Nitzan}

\affiliation{School of Chemistry, Tel Aviv University, Tel Aviv,
69978, Israel}

\date{\today}

\begin{abstract}
Rectification of heat transfer in nanodevices can be realized by
combining the system inherent anharmonicity with structural
asymmetry. we analyze this phenomenon within the simplest
anharmonic system -a spin-boson nanojunction model. We consider
two variants of the model that yield, for the first time,
analytical solutions: a linear separable model in which the heat
reservoirs contribute additively, and a non-separable model
suitable for a stronger system-bath interaction. Both models show
asymmetric (rectifying) heat conduction when the couplings to the
heat reservoirs are different.
\end{abstract}


\maketitle

The heat conduction properties of nanojunctions attract attention for
two reasons. First, heating in nanoconductors, a crucial issue for
their operation and stability, is determined by both heat release
and conduction in such systems. Secondly, as with electronic
conduction, the restrictive geometry raises fundamental questions
concerning the relationship between transport processes in
microscopic systems and their macroscopic counterparts. Indeed
thermal transport properties of nanowires can be very different
from the corresponding bulk properties as is demonstrated by the
recent confirmation \cite{Schwab} of the prediction \cite{Rego}
that the low temperature ballistic phonon conductance in a 1
dimensional quantum wire is characterized by a universal quantum
unit. Also of considerable interest are studies that confront the
macroscopic Fourier law, $J=-K\nabla T$ , a linear relationship
between the heat current $J$ and the temperature gradient $\nabla
T$ that defines the thermal conductance $K$, with heat transport
on the microscopic scale \cite{Rieder,Talkner,Casher, Lepri,
Morkoss, Hu}. Harmonic chains were repeatedly discussed
theoretically in these contexts and considerable experimental
progress was also made \cite{Schwab,Cahill,Shi}. For reviews
see Refs. \cite{Ford,Livi}.

An intriguing mode of behavior often addressed in the study of
transport devices is current rectification, allowing larger
conduction in one direction than in the opposite one when driven
far enough from equilibrium. Such phenomena were extensively
studied for electronic conduction in molecular junctions, but much
less so for thermal nano-conductors. For a harmonic thermal
conductor connecting (by linear coupling) two (left (L), right
(R)) harmonic thermal reservoirs that are maintained at
equilibrium with the temperatures $T_L$ and $T_R$, respectively,
heat transfer is a ballistic process and the heat current $J$ can
be recast into a Landauer type expression \cite{Rego,Ciraci,Segal}
\beq J(\omega)=\int \mathcal{T}(\omega)\left[ n_L(\omega) -
n_R(\omega) \right] \omega d\omega,
 \label{eq:Landauer}
 \eeq
 where $\mathcal{T}(\omega)$ is the transmission coefficient for phonons of frequency $\omega$
and $n_{K}(\omega)=\left( e^{\beta_{K} \omega} -1 \right)^{-1}$;
$\beta_K=(k_{B}T)^{-1}$; $K=L,R$ ($\hbar \equiv 1$) are
Bose-Einstein distribution functions characterizing the
reservoirs. Obviously, this expression is symmetric to
interchanging the reservoirs temperatures and cannot show
rectifying behavior irrespective of any asymmetry in the system
structure.

\begin{figure}[htbp]
\vspace{3mm}
\hspace{-2mm}
 {\hbox{\epsfxsize=70mm \epsffile{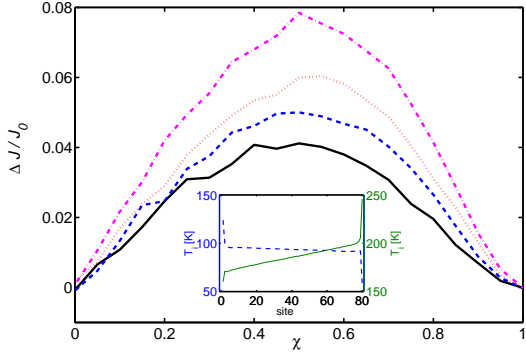}}}
\vspace{-3mm} \caption{ The asymmetry in the thermal conduction
plotted against $\chi$ for a classical $N$-atom chain. 
The parameters used, $D=3.8/ \nu^2$ eV, $\alpha=1.88 \nu$
\AA$^{-1}$, $x_{eq}$=1.538 \AA$ $  and $m=12/6.02$ $10^{-23}$ g,
are based on a standard model for the carbon-carbon force field in
alkanes \cite{Lifson}, for which $\nu$=1. Here we increase the
system anharmonicity by taking $\nu$=6. Full, dashed, dotted and
dashed-dotted lines correspond to $N$=10, 20, 40 and 80,
respectively, with $\gamma$=50 ps$^{-1}$, $T_h$ = 300 K and $T_c$=
0 K. (Inset) The temperature profile for the $N$=80, $\chi$=0.5
case with $T_L=T_c$; $T_R=T_h$ (full), $T_L=T_h$ ; $T_R=T_c$
(dashed). }
 \label{Fig1}
\end{figure}

In contrast, Terraneo et al \cite{Terraneo} have shown numerically
that rectifying behavior is obtained by replacing the interior
part of a classical harmonic chain by an anharmonic segment. An
example of a similar behavior by a somewhat simpler model is shown
in Fig. \ref{Fig1}. The model is defined by the $N$-particle
Hamiltonian
\begin{eqnarray}
 H= (2m)^{-1}\sum_{i=1}^{N} p_i^2   + \sum_{i=1}^{N-1} D
\left( e^{ -\alpha(x_{i+1} -x_i-x_{eq} ) }-1 \right)^2 +
\nonumber\\
 D \left( e^{
-\alpha(x_{1} -a ) }-1 \right)^2 + D \left( e^{ -\alpha(b-x_{N}
 ) }-1 \right)^2
 \end{eqnarray}
supplemented by damping and noise terms operating on particle 1
and $N$ to simulate the effect of two thermal baths. The equations
of motions are $\ddot{x}_{i}=-(1/m){\partial H} /{\partial x_i} -
\left( \gamma_L \dot{x}_1  -F_L(t) \right)\delta_{i,1} - \left(
\gamma_R \dot{x}_N -F_R(t) \right) \delta_{i,N}$. In these
equations $a$, $b$, $x_{eq}$, $\gamma$ and $m$ are constants and
$F_K(t)$, $K=L,R$ are Gaussian random forces that satisfy $\langle
F_K(t)F_K(0)\rangle =2 \gamma_K k_B T_K \delta(t)/m$. We take
$\gamma_L=\gamma(1-\chi)$ and $\gamma_R=\gamma(1+\chi)$ ,$|\chi|
\leq 1$, and study the ratio between $\Delta J \equiv J(T_L=T_h;
T_R=T_c)+J(T_L=T_c;T_R=T_h)$ and $J_0\equiv|J(\chi=0)|$, where
$T_c$ and $T_h$ denote low and high temperatures and where the
heat current $J$ is calculated as the average over sites, at
steady state, of $ J_i=\langle -\dot{x}_i \left( \partial
H_{i+1,i}/ \partial x_i\right) \rangle $ with $H_{i+1,i}=D \left(
e^{-\alpha(x_{i+1} - x_{i}-x_{eq})}-1   \right)^2$.
Fig. 1 shows that the intrinsic non-linearity of this model is
enough to induce asymmetry in the thermal conduction of the
asymmetrically coupled ($\gamma_L\neq \gamma_R$ ) bridge.

Clearly, in addition to structural asymmetry, non-linear
interactions are essential for rectifying behavior. In this paper
we examine the rectifying properties of the simplest non-linear
heat conductor: a two level system (TLS). The model investigated
is a generalization of the spin-boson model \cite{TLS} in which
the TLS is coupled to two equilibrium boson baths maintained at
different temperatures. We study two variants of the model and
show that if asymmetry is built into either one by employing, e.g.
different spin-boson coupling strengths for the two baths, thermal
rectification naturally sets in.
%

The first variant of our spin-boson model is defined by
the Hamiltonian
\beq H=E_0 |0\rangle \langle0| + E_1 |1\rangle \langle1| +H_B +H_{MB} \label{eq:Eq3} \eeq
\beq H_B=H_L+H_R; \   \ H_K=\sum_{j\in K}{\omega_j a_j^\dagger
a_j} ;\  \ K=L,R \label{eq:Eq4}
\eeq
\beq
 H_{MB}=B|0\rangle \langle1| +B^{\dagger}|1\rangle \langle0|; \ \  B=B_L+B_R,
\label{eq:Eq5}
\eeq
where $a_j^{\dagger}$, $a_j$ are boson creation and annihilation
operators
 associated with the phonon modes of the harmonic baths and $B_K$
are bath operators. For a linear coupling model
\beq B_K= \sum_{ j \in K}{\bar{\alpha}_j x_j}; \ \
x_j=(2\omega_j)^{-1/2}({a}_j^{\dagger}+a_j) ;\ \ K=L,R.
\label{eq:Eq6} \eeq

An important attribute of this model is the mutual independence of
the transport processes at the two system-bath interfaces. It is
convenient to regard the model (\ref{eq:Eq3})-(\ref{eq:Eq6}) as a
special case of an $N$ equally spaced states system with the
nearest-neighbor coupling $H_{MB}=\sum_{n=1}^{N-1} \sqrt{n} \left(
B|n-1\rangle \langle n| +B^{\dagger} |n\rangle  \langle n-1|
\right)$. In particular the $N\rightarrow \infty$ limit
corresponds to a harmonic oscillator bridge connecting the baths.
Eq.~(\ref{eq:Eq6}) corresponds in the latter case to the bilinear
coupling model for the oscillator-baths
interactions, $H_{MB}= \sum_{j \in K} \alpha_j x_j x$, 
where $x$ is the coordinate of the bridge oscillator and
${\alpha_j}=\bar{\alpha}_j(2m\omega_0)^{1/2}$, where $m$ and
$\omega_0=E_1-E_0$ are the oscillator mass and frequency,
respectively.

The reduced dynamics of the $N$-level system can be derived
following standard procedures, e.g. the Redfield approximation
\cite{Redfield} for the weak system-baths coupling limit. In the
limit of fast dephasing the resulting kinetic equations for the
state probabilities are
\bea
\dot P_n=- \left( nk_d+(n+1)k_u X_n \right)P_n +nk_uP_{n-1}
\nonumber\\
+(n+1) k_d  X_n P_{n+1};\ \ P_{-1}=0, \label{eq:Eq7}
\eea
where $X_n=\delta_{n,0}$ for the two level ($n$=0,1) system and
$X_n$=1 for the harmonic oscillator ($n$ = 0,..,$\infty$ ) case,
and where $k_d= \int_{-\infty}^{\infty} d\tau e^{i \omega_0 \tau}
\langle B^{\dagger}(\tau) B(0) \rangle $ and  $k_u=
\int_{-\infty}^{\infty} d \tau e^{-i \omega_0 \tau} \langle
B(\tau) B^{\dagger}(0) \rangle$. The average is over the baths
thermal distributions, irrespective of the fact that it may
involve two distributions of different temperatures \cite{linear}.
Specifying to the linear coupling model, and assuming no
correlation between the thermal baths leads to the rates
\bea k_d=k_L+k_R; \ \  k_u=k_L e^{-\beta_L \omega_0}+k_R
e^{-\beta_R \omega_0},
\label{eq:Eq8}
\eea
 with
\beq k_L=\Gamma_L(\omega_0) (1+n_L(\omega_0)) ; \ \
k_R=\Gamma_R(\omega_0) (1+n_R(\omega_0)),
\label{eq:Eq9} \eeq
\beq \Gamma_K(\omega)= \frac{\pi}{2m\omega^2} \sum_{j \in K}
\alpha_j^2\delta(\omega-\omega_j); \ \  K=L,R. \label{eq:Eq10}
\eeq
The heat conduction properties of this model are obtained from the
steady state solution of Eqs.~(\ref{eq:Eq7}) with the rates given
in Eqs.~(\ref{eq:Eq8})-(\ref{eq:Eq10}). For the harmonic
model ($N \rightarrow \infty $), putting $\dot{P}_n=0$, 
and searching a solution of the form $P_n \propto y^n$ we get a
quadratic equation for $y$ whose physically acceptable solution is
\beq
y=\frac{k_L e^{-\beta_L \omega_0} + k_R e^{-\beta_R \omega_0}
}{k_L+k_R}.
\label{eq:Eq11}
\eeq
This leads to the normalized state populations $P_n=y^n(1-y)$.
The steady-state heat flux is
obtained from
\bea J=\omega_0 \sum_{n=1}^{\infty} n \left( k_R P_n -k_RP_{n-1}
e^{-\beta_R \omega_0} \right)
\label{eq:Eq13}
\eea
where positive sign indicates current going from left to right \cite{flux}.
Using Eqs.~(\ref{eq:Eq9}) and (\ref{eq:Eq11}) we find
\beq J=\omega_0 \frac{\Gamma_L \Gamma_R}{\Gamma_L+\Gamma_R} \left(
n_L-n_R \right).
\label{eq:Eq14}
\eeq
This is a special case (with $\mathcal{T}(\omega)= \Gamma_L
\Gamma_R(\Gamma_L+\Gamma_R)^{-1} \delta(\omega-\omega_0) $
consistent with our resonance energy transfer assumption)
\cite{general} of Eq.~(\ref{eq:Landauer}). Obviously no rectifying
behavior is obtained in this limit.

Next consider the two levels case, $N$=2. The two steady-state
equations obtained from (\ref{eq:Eq7}) yield
\beq
P_1=\frac{k_L e^{-\beta_L \omega_0}  + k_R e^{-\beta_R \omega_0} }
{k_L(1+ e^{-\beta_L \omega_0}) + k_R(1+e^{-\beta_R \omega_0})}; \ \
P_1=1-P_0,
\label{eq:Eq15}
\eeq
and the analog of Eq.~(\ref{eq:Eq13}) is
\beq J=\omega_0 k_R\left(  P_1 - P_0 e^{-\beta_R \omega_0}
\right).
\label{eq:Eq16} \eeq
Using this with Eq.~(\ref{eq:Eq9}), leads to
\beq J=\omega_0 \frac{\Gamma_L \Gamma_R \left( n_L-n_R
\right)}{\Gamma_L \left( 1+ 2n_L \right) + \Gamma_R \left( 1+
2n_R \right)},
\label{eq:Eq17}
\eeq
which does have rectifying behavior. Indeed, defining the
asymmetry parameter $\chi$ such that $\Gamma_L=\Gamma(1-\chi)$
 ;   $\Gamma_R=\Gamma(1+\chi)$ with $-1 \leq \chi \leq 1$ we find
\bea \Delta J \equiv J(T_L=T_h; T_R=T_c) +J(T_L=T_c; T_R=T_h)
\nonumber\\
=\frac{\omega_0 \Gamma \chi (1-\chi^2)
(n_L-n_R)^2}{(1+n_L+n_R)^2-\chi^2(n_L-n_R)^2}.
 \label{eq:Eq18}
 \eea
Eq.~(\ref{eq:Eq18}) implies that for small $\Delta T=T_L-T_R$, $
|\Delta J|$ grows like $\Delta T^2$. Furthermore, noting that
$sign(\Delta J)=sign(\chi)$ it follows from (\ref{eq:Eq18}) that
the current is larger when the bridge links more strongly to
colder reservoir than when it links more strongly to the hotter
one. Figure 2 shows an example of this behavior.
\begin{figure}[htbp]
\vspace{3mm} \hspace{-2mm}
 {\hbox{\epsfxsize=70mm \epsffile{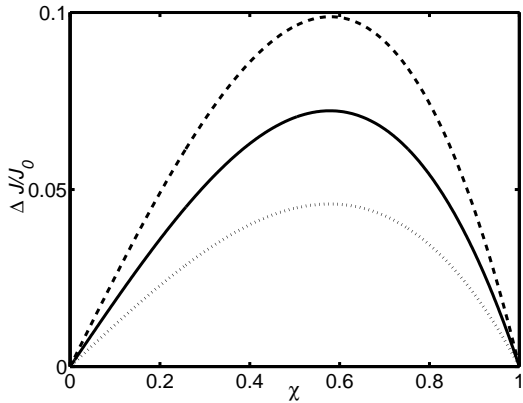}}}
\vspace{-3mm} \caption{ Heat rectification by a TLS bridge in the
linear coupling model, Eqs.~(\ref{eq:Eq3})-(\ref{eq:Eq18}). The
ratio $\Delta J / J_0$ (with $J_0 \equiv |J(\chi=0)|$ ) is plotted
against the asymmetry parameter $\chi$ for several two-level
spacings: $\omega_0$=0.025 eV (dashed); $\omega_0$=0.05 eV (full);
$\omega_0$=0.075 eV (dotted). The temperatures are $T_h$=400 K,
$T_c$=300 K.}
 \label{Fig2}
\end{figure}
%


Next we consider another variant of the two-bath spin-boson model, taking the Hamiltonian to be
\begin{eqnarray}
 H= E_0|0\rangle \langle 0|+ E_1|1\rangle \langle 1| + V_{0,1}|0\rangle \langle 1|
 + V_{1,0}|1\rangle \langle 0|
\nonumber\\
 +\sum_{j \in L,R} \omega_j a_j^{\dagger}a_j
+\sum_{j \in L,R}x_j \left( \alpha_{0,j}|0\rangle \langle 0| +
 \alpha_{1,j}|1\rangle \langle 1| \right) .
 \label{eq:Eq19}
 \end{eqnarray}
The $L$ and $R$ boson baths are again maintained at different
temperatures $T_L$ and $T_R$. When $T_L=T_R$, Eq.~(\ref{eq:Eq19})
represents a standard spin-boson Hamiltonian used, e.g, in the
electron transfer problem. Using the small polaron transformation
\cite{Mahan}, $\tilde{H}= UHU^{-1}$, leads to
\bea \tilde{H}&=&E_0|0\rangle \langle 0|+ E_1|1\rangle \langle 1|
+ V_{0,1}|0\rangle  \langle 1|e^{i \Omega}
 + V_{1,0}|1\rangle \langle 0|e^{-i\Omega}
\nonumber\\
& +& \sum_{j \in L,R} \omega_j
 a_j^{\dagger}a_j+H_{shift},
 \label{eq:Eq20}
 \eea
where $U=U_0 U_1$, 
 $U_n=exp(-i\Omega_n|n\rangle \langle n|)$, ($n$=0,1),
$\Omega_n=\Omega_n^L+ \Omega_n^R$, $\Omega_n^K=i \sum_{j \in K}
\lambda_{n,j}\left( a_j^{\dagger} -a_j \right)$ ($K=L,R$),
$\lambda_{n,j}=(2\omega_j^3)^{-1/2}\alpha_{n,j}$ and
$\Omega=\Omega_1-\Omega_0$. The term
$H_{shift}=-(1/2)\sum_{j}\omega_j^{-2} \left( \alpha_{0,j}^2
|0\rangle \langle 0| + \alpha_{1,j}^2 |1\rangle \langle 1|\right)$
may be henceforth incorporated into the zero order energies. The
Hamiltonian (\ref{eq:Eq20}) is similar to that defined in Eqs.
(\ref{eq:Eq3})-(\ref{eq:Eq5}), except that the system-baths
couplings appear as multiplicative rather than additive factors in
the interaction term, implying non-separable transport at the two
contacts. The dynamics is still readily handled. For small $V$
(the "non-adiabatic limit") the Hamiltonian (\ref{eq:Eq20}) leads
again to the rate equation (\ref{eq:Eq7}) with
\beq k_d=|V_{0,1}|^2 C(\omega_0);\ \  k_u=|V_{0,1}|^2
C(-\omega_0), \label{eq:Eq21}
 \eeq
where $C(\omega_0)=\int_{-\infty}^{\infty}dt e^{i\omega_0 t}
\tilde{C}(t)$ and
\bea \tilde{C}(t)&=& \left<  e^{i\Omega(t)}e^{-i\Omega(0)} \right>
=\left<  e^{i(\Omega_1^L(t) - \Omega_0^L(t) )} e^{-i( \Omega_1^L
-\Omega_0^L) } \right>_L
\nonumber\\
& \times & \left<  e^{i(\Omega_1^R(t) - \Omega_0^R(t) )} e^{-i(
\Omega_1^R -\Omega_0^R) } \right>_R.
 \label{eq:Eq22}
 \eea
This may be evaluated explicitly to give
\beq \tilde{C}(t)=\tilde{C}_L(t)\tilde{C}_R(t); \ \
\tilde{C}_K(t)=exp(-\phi_K(t)) \ \ , \label{eq:Eq23}
 \eeq
\bea
\phi_K(t)=\sum_{j \in K}( \lambda_{1,j}- \lambda_{0,j})^2 [
\left( 1+2n_K(\omega_j) \right)
 \nonumber\\
  - \left(
1+n_K(\omega_j) \right)e^{-i \omega_j t} -
n_K(\omega_j)e^{i\omega_j t} ]. \label{eq:Eq24}
 \eea
Explicit expressions may be obtained using the short time
approximation (valid for  $\sum_{j\in K}(\lambda_{1,j}-
\lambda_{0,j})^2 \gg 1$ and/or at high temperature) whereupon
$\phi(t)$ is expanded in powers of $t$ keeping terms up to order
$t^2$. This leads to
\beq C(\omega_0)=\sqrt{ \frac{2 \pi}{(D_L^2+D_R^2)}} exp\left[
\frac{-(\omega_0-E_M^L-E_M^R)^2}{2(D_L^2+D_R^2)} \right],
\label{eq:Eq25}
 \eeq
where $E_M^K=\sum_{j \in K} (\lambda_{1,j}- \lambda_{0,j} )^2
\omega_j$, $D_K^2=\sum_{j \in K} (\lambda_{1,j}- \lambda_{0,j} )^2
\omega_j^2\left( 2n_K(\omega_j)+1 \right) \stackrel {\omega/k_B
\rightarrow 0}  {\longrightarrow} 2k_BT_KE_M^K$.
Eqs.~(\ref{eq:Eq21})-(\ref{eq:Eq25}) provide an extension of the
Marcus non-adiabatic rate expressions \cite{Marcus} to the case of
two reservoirs maintained at different temperatures. $E_M^L$ and
$E_M^R$ are the corresponding reorganization energies.

\begin{figure}[htbp]
\vspace{3mm} \hspace{-2mm}
 {\hbox{\epsfxsize=65mm \epsffile{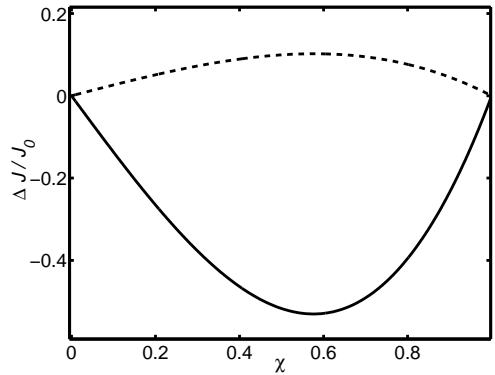}}}
\vspace{-3mm} \caption{Heat rectification of a TLS bridge with a
non-separable coupling, Eqs. (\ref{eq:Eq19})-(\ref{eq:Eq29}).
$\Delta J/ J_0$ is plotted against $\chi$ for $\omega_0$=0.025 eV
and for $E_M$=0.012 eV (dashed), $E_M$ =0.38 eV (full). The
temperatures are $T_h$=400 K, $T_c$=300 K. }
 \label{Fig3}
\end{figure}

Consider now the steady state heat current. The non-separability
of the system-bath couplings makes the procedure that leads to
Eq.~(\ref{eq:Eq16}) unusable. Instead note that $C_L(\omega_0)$
and $C_R(\omega_0)$ are the rates affected by each thermal
reservoir separately and that, from (\ref{eq:Eq23}),
$C(\omega_0)=\int_{-\infty}^{\infty}d\omega
C_L(\omega_0-\omega)C_R(\omega)$. The process $|1 \rangle
\rightarrow |0 \rangle $ in which the TLS looses energy $\omega_0$
can be therefore viewed as a combination of processes in which the
system gives energy $\omega$ (or gains it if $\omega<0$ ) to the
right bath and energy $\omega_0-\omega $ to the left one, with
probability $C_L(\omega_0-\omega)C_R(\omega)$. A similar analysis
applies to the process $|0\rangle  \rightarrow |1\rangle $. The
heat flux calculated as the energy transferred per unit time into
the right bath is therefore \cite{flux}
\begin{eqnarray}
J=|V_{0,1}|^2 \int_{-\infty}^{\infty} d\omega \omega [
C_R(\omega)C_L(\omega_0-\omega )P_1
\nonumber\\
 -C_R(-\omega)C_L(-\omega_0+\omega)P_0 ],
 \label{eq:Eq28}
 \end{eqnarray}
where $P_0=C(\omega_0)/\left( C(\omega_0) + C(-\omega_0) \right)$
and $P_1=1-P_0$ are the steady state probabilities that the system
is in state 0 or 1, respectively. In the short time approximation
$C(\omega)$ takes the form
$C_K(\omega)=(D_K^2)^{-1/2}exp\left[-(\omega-E_M^K)^2/2D_K^2
\right]$. It is convenient also to take $E_M^L=E_M(1-\chi)$;
$E_M^R=E_M(1+\chi)$ ($|\chi| \leq 1$), which implies
$D_L^2+D_R^2=2k_BE_M(T_S-\chi\Delta T)$ where 
$T_S=T_L+T_R$. Using these relationships in (\ref{eq:Eq28}) leads
to
\bea J=\frac{2\sqrt{\pi} |V_{0,1}|^2 (1-\chi^2) E_M^2 k_B  \Delta
T}{\left( k_BE_M(T_S-\chi \Delta T) \right)^{3/2}}
 \frac{e^{- \frac{(\omega_0-2E_M)^2}{4k_BE_M(T_s-\chi \Delta
T)}}} {1+e^{2\omega_0/k_B(T_S-\chi \Delta T)}}.
 \label{eq:Eq29}
 \eea
Eq. (\ref{eq:Eq29}) again implies asymmetric heat conduction
provided symmetry is broken by taking $\chi \neq 0$. This is shown
in Fig. \ref{Fig3} where $\Delta J/J_0$ is displayed against
$\chi$. It is seen that the heat conduction asymmetry can be quite
large, with its magnitude and sign strongly dependent on system
parameters. When $E_M \gg \omega_0 $ the heat flux is dominated by
the term $ e^{-(\omega_0-2E_M)^2/4k_BE_M(T_s-\chi \Delta T)}$ that
is bigger when $\Delta T$ is negative than when it is positive,
hence the negative asymmetry in $\Delta J$. The same behavior is
seen in the opposite limit, $E_M \ll \omega_0$. However, when
$2E_M \approx \omega_0$ and $\omega_0 \approx k_BT$, $J$ is
dominated by the term $\left( k_BE_M(T_S-\chi \Delta T)
\right)^{-3/2}$, implying positive asymmetry as seen in Fig.
\ref{Fig3}.
%

In Summary, while rectification of electronic current in molecular
junctions is well known, heat flux rectification is a novel
concept. Asymmetric anharmonic chains have this property. We have
presented a simple heat rectifying model where anharmonicity stems
from the dynamics of a two-levels system and asymmetry is
introduced by different interaction strengths with the thermal
baths. We have considered two cases that yield analytical
solutions: a separable model with additive interactions with the
bridge, and a non separable model. For both, the calculated heat
current shows diode like behavior that depends on the junction
characteristics.

Asymmetric coupling to the two thermal reservoirs can be affected
by different chemical bonding or by using phonon reservoirs with
different Debye temperatures. Alternatively, in a more realistic
molecular model, this may result from asymmetric spatial
organization of the molecular vibrational states.

Heat rectification will be very useful in nanodevices, where
efficient heat transfer away from the conductor center is crucial
for proper functionality and stability. Similarly, directed energy
flow in biomolecules such as proteins \cite{Leitner2} may play a
role in controlling conformational dynamics. From the theory
perspective, the efficiency of rectification and its possible
optimization 
are issues for future considerations.

\begin{acknowledgments}
This research was supported by the Israel National Science
Foundation and by the U.S. - Israel Binational Science Foundation.
\end{acknowledgments}



\begin{thebibliography}{99}
\bibitem{flux}
The heat flux can be equivalently calculated at the left contact.
\end{thebibliography}
\end{document}